# High temperature superconductivity ($T_c$ onset at $34K$) in the high pressure orthorhombic phase of $FeSe$

**G. Garbarino**
European Synchrotron Radiation Facility (ESRF), 6 Rue Jules Horowitz, BP 220, F-38043 Grenoble Cedex France EU

**A. Sow, P. Lejay, A. Sulpice, P. Toulemonde**
Institut NEEL, CNRS & Université Joseph Fourier, 25 Avenue des Martyrs, BP166, F-38042 Grenoble Cedex 9 France EU

**M. Mezouar**
European Synchrotron Radiation Facility (ESRF), 6 Rue Jules Horowitz, BP 220, F-38043 Grenoble Cedex France EU

**and M. Núñez-Regueiro**
Institut NEEL, CNRS & Université Joseph Fourier, 25 Avenue des Martyrs, BP166, F-38042 Grenoble Cedex 9 France EU

We have studied the structural and superconducting properties of $\beta - FeSe$ under pressures up to $26 GPa$ using synchrotron radiation and diamond anvil cells. The bulk modulus of the tetragonal phase is $28.5(3) GPa$, much smaller than the rest of $Fe$ based superconductors. At $12 GPa$ we observe a phase transition from the tetragonal to an orthorhombic symmetry. The high pressure orthorhombic phase has a higher $T_c$ reaching $34 K$ at $22 GPa$.

email: garbarin@esrf.fr

74.70.-b, 61.50.Ks

The discovery of superconductivity[1,2,3,4] with critical temperatures between 30K and 55 K in the layered iron arsenides has induced a wide scope exploration of phases with iron in a similar environment in the hope of finding new superconducting compounds and going beyond these high superconducting transition temperatures ($T_c$). In particular, the report[5] of superconductivity in the *PbO* structure of *FeSe* is tantalizing, as strong pressure effects at relatively low pressures seem[6,7,8] to strongly increase $T_c$ up to around $25K$ for the onset of the transition. On the other hand, partial replacement with tellurium also noticeably increases $T_c$[9]. However, superconductivity seems to coexist with magnetically ordered states accompanied by lattice distortions [10] that are much more stable than superconductivity[11]. In this letter, we have studied the evolution of both superconductivity and structure under pressure. We find that in our samples an orthorhombic high pressure phase develops above $12GPa$ that presents superconductivity above 34K.

Polycrystalline samples were synthesized from high quality starting materials: Fe $4N$ and Se $5N$ in molar ration of $1:0.82$. The reactants were weighted, mixed and homogenized in an agate mortar in a dry box under purified argon atmosphere. The mixture was packed and sealed under vacuum in a quartz tube and calcined at $700C$ during 48h with intermediate grindings. The polycrystalline sample was checked using X-ray powder diffraction technique. The pattern was indexed according to the tetragonal *FeSe* with $P4/nmm$ space group. Small peaks of the hexagonal $\alpha-FeSe$ (*NiAs*-type) were observed as an impurity phase.

The angle dispersive X-ray diffraction studies on *FeSe* powder samples were performed at the ID27 high-pressure beamline of the European Synchrotron Radiation Facility using monochromatic radiation ($\lambda=0.26473\text{Å}$) and diamond anvil cells with $300\mu m$ cullet diamonds. The transmitting media used was Helium and the pressure was determined using the equation state of copper[12], introduced next to the sample. The structural studies have been done at ambient temperature. The diffraction patterns were collected with a CCD camera and the intensity vs. *2Theta* patterns were obtained using the fit2d software[13]. A complete Rietveld refinement was done with the GSAS-EXPGUI package[14], all the estimated standard deviations were multiplied by a factor of 5 due to the underestimation of the GSAS calculation[15].

The electrical resistance measurements were performed using a Keithley 220 source meter and a Keithley 2182 nanovoltmeter. Pressure measurements, $2-22 GPa$ (between $4.2K$ and $300K$), were done in a sintered diamond Bridgman anvil apparatus using a pyrophillite gasket and two steatite disks as the pressure medium[16].

The magnetization (M) data was measured in the so-called zero field cooling process. The sample was cooled down to 2K with in a zero field. Then, the 50Oe magnetic field was applied and M was measured at increasing temperature up to room temperature.

From the diffraction data at ambient pressure condition, we refined the lattice parameters, peak profile, occupancy and position $z$ of the Se atom in $(\frac{1}{4}, \frac{1}{4}, z)$, being the Fe in a high symmetry Wyckoff position $(\frac{1}{4}, -\frac{1}{4}, 0)$. We obtained $a = 3.7699(2)Å$, $c = 5.5184(6)Å$ and $V = 78.429(8)Å$ for the lattice parameters and volume of the tetragonal phase, consequently $c/a = 1.464(1)$, while the superconducting critical temperature was $T_c = 7K$ from the magnetic measurement (Fig. 3(a)). From this results, and considering the study presented in Ref 17 we can conclude that our sample correspond to an stoichiometric composition $FeSe$, that is consistent with the Se occupancy refined from the ambient pressure X ray data of $1.04(5)$.

On Fig.1(a) we show the X-ray diffraction pattern and the Rietveld refinement at 6.6GPa. The refinement was done considering both the tetragonal $\beta - FeSe$ and the mentioned hexagonal impurity phase $\alpha - FeSe$ phase, with space groups $P4/nmm$ and $P6_3/mmc$, respectively. The pressure effects on the diffraction patterns can be observed in fig.1 (b). Clearly, at 12.5GPa there is a phase transition from a tetragonal towards an orthorhombic (T-O) phase. Although it was not straightforward to determine the space group of this phase, considering group-subgroup operation, we can propose a structure of the $Pbnm$ type. It is also important to mention that we also observed the phase transition in the hexagonal phase at $8.5GPa$ already reported in the literature[18].

By inspection of the lattice parameters of the tetragonal phase, determined by Rieteveld refinement, we observe a smooth variation over the full range to $12GPa$. The lattice parameters $a, c$ and the volume $V$ of the $\beta - FeSe$ phase are 6, 36 and 44% smaller than in the case of LaFeAsO[19] and 0.1%, 13% and 15% in the case of LiFeAs[20], respectively. This reduction may explain the low critical pressure of the structural phase transition.

The pressure evolution of the lattice parameters $a$ and $c$, the normalized ratio $c/a$ and the volume $V$ of the tetragonal phase, are presented in fig. 2(a) and (b), respectively. The $c/a$ ratio decreases down to 1.39 at $12GPa$ meaning a larger compression of the structure along the stacking direction of the layers compared with the basal plane and it is consistent with the higher atom density in the $ab$ plane.

From a fit of the cell volume, $V$, with pressure, $P$ up to $12GPa$, with a third order Murnaghan Equation of State (EoS) $V = V_0 \left(1 + K_0' P / K_0 \right)^{-1/K_0'}$, we obtained a bulk modulus $K_0 = 28.9(3) GPa$ and $K_0' = 6.1(1)$. This is the lowest $K_0$ value reported up to now in the iron based superconductors family. Such a small value can explain the low critical pressure $(12.5 GPa)$ for the T-O transformation. A similar value for the bulk modulus was obtained for this compound in recent high pressure neutron diffraction studies at $190 K$ [21].

From the magnetic measurement at ambient pressure, we can clearly observe the existence of two anomalies. The first at 125K (named $T^*$) can be associated with the spin density wave transition, while the second one at 7K evidences the appearance of the superconductivity (see fig. 3 (a)). In the same figure the temperature dependence of the resistance at $6 GPa$ is presented, where both the anomaly at $T^*$ and the superconductivity are marked by an arrow.

In the fig. 3 (b) the temperature dependence of the normalized resistance is shown for various applied pressures up to 22GPa. The pressure dependence of the resistance at 50K (inset of Fig. 3 (a)) shows a kink at 13GPa. This anomaly can be associated to the T-O phase transition determined from the high pressure diffraction studies.

From the Rietveld refinement of X-ray data, we obtained the pressure dependence of the characteristics distances between $Se$ and $Fe$ atoms (see fig. 3 (c)) and the $Se-Fe-Se$ angles (see fig. 3 (d)). The Se height in the structure does not show an evident variation while the pressure increases; it remains fixed at 1.476(15)Å from the basal plane. The interlayer $Se-Se$ bond length ($Se(1)-Se(2)$ in fig. 3 (c)) has a higher compression than the intralayer $Se-Se$ distance ($Se(1)-Se(1)$ in fig. 3 (c)). Therefore, the separation between layers (labeled "$BetweenLayers$" in fig. 3(c)) decreases abruptly, reaching a value below 2Å at 12GPa. In contrast, the $Fe-Se$ bond length ($Fe(1)-Se(1)$ in fig. 3(c)) and the thickness of the $FeSe$ layers ($LayerFeSe$ in fig. 3(c)) show a slight decrease at low pressure (P<6GPa) followed by a constant behavior up to 12GPa. The $Se-Fe-Se$ angles at ambient pressure are similar to the ones obtained in Ref 10 and correspond to a deviation from the ideal value for the non-distorted $FeAs_4$ tetrahedron, 109.47°. For pressure below 1GPa, the difference between the two angles increases rapidly inducing a bigger distortion in the structure, and then it remains constant up to 6GPa. Further increase in pressure develops again the distortion of the structure.

The pressure dependence of $T_{c-onset}$, $T_{c-mid}$ (onset and middle point of the superconducting transition from resistivity measurements, respectively) and $T^*$ are shown in Fig. 4. We also present in this figure with two different background colors the stability pressure range of the tetragonal and orthorhombic phase.

It is clear that the structural phase transition at 12.5GPa induces an abrupt change of the superconducting temperature pressure dependence. Thus, the orthorhombic phase has a higher $T_c$ that reaches its maximum value of 34K at the highest applied pressure. In contrast, the $T^*$ transition shows a monotonic increase up to 15GPa followed by a reduction that could be fitted with a quadratic polynomial.

For the low pressure variation ($P < 12GPa$) of $T_{c-onset}$ and $T_{c-mid}$, we obtained 1.4K/GPa and 0.8K/GPa, respectively. These values are smaller that the ones reported in Ref. 7 (~4K/GPa) and Ref. 8 (5K/GPa).

During the preparation process of this publication, two other papers[22,23] have been published combining high pressure x-ray diffraction, resistivity (Ref 23) and Mossbauer studies (Ref 22) on $FeSe$.

In Ref. 22, the authors showed a wide pressure range ($7GPa - 35GPa$) phase transition from the tetragonal to the hexagonal phase. The $T_c$ increases with pressure from $8.5K$ with a slope $dT_c/dP = 3.2K/GPa$ up to $8.9GPa$ where it shows a maximum of $36K$ and then it decreases for further increasing pressure. In Ref. 23, high pressure X-ray diffraction measurements at low temperature ($T = 16K$) show a phase transition from the orthorhombic $Cmma$ phase to the hexagonal $NiAs-type$ at $9GPa$. From the high pressure resistance measurements, the authors show a $T_c(P)$ dependence similar to the one presented in Ref. 22, with a maximum of $37K$ at $7GPa$. For higher pressures, $T_c$ decreases reaching $6K$ at $14GPa$.

Our high pressure structural and resistance results show important discrepancies with these two recent results. First, the tetragonal phase transform to an orthorhombic phase, where the $T_c$ reaches its maximum at $34K$ and $22GPa$. Second, the slope $dT_c/dP$ is smaller and we did not observe a decrease of $T_{c-Onset}$ with pressure. Finally, we show no evidence of a high pressure hexagonal $NiAs-type$ phase.

Before a detailed analysis, our structural results show lattice parameters at room pressure and temperature compatible with the ones reported in Ref 22, but 0.1% smaller than in the case of Ref 23. For the bulk modulus, we obtain similar values to the one in Ref 23, but it is smaller in the case of Ref 22, considering that $c/a = 1.414$ is reached at $6.6GPa$ in our measurement instead of $1.5GPa$. This huge variation in compressibility values could explain the discrepancy in the superconducting behavior under pressure.

To try to elucidate other reason for the large differences we discuss the possible effect of the pressure transmitting media. As it has been mentioned before, the structural study was done using Helium as media. Taking into account the small atomic radii of the noble gas, one can speculate about He intercalation during the compression process considering the layered crystal structure of the $FeSe$. Three important aspects must be taken into account in order to discard any effect of the pressure media. First, the resistance experiment was done using solid steatite as media where no possible intercalation is expected and even there is an excellent agreement in the determination of the critical pressure of the T-O transformation from the kink in the resistance at

50K, and in the pressure dependence of the superconducting critical temperature. Second, the T-O phase transition was completely reversible during decompression in the diffraction study. Finally, the fact that the neutron diffraction results and the data presented in Ref. 23 (where Daphne$^{TM}$ oil was used as pressure media) show the same bulk modulus helps discarding the He intercalation hypothesis. Using this three arguments we can also discard any oxygen doping or moisture effect on the sample, considering that in this case the cell parameters would be increased and the $T_c$ drastically modified. The $T_c$ measurements in Ref. 7-8, 22 and 23 were performed using hydrostatic conditions, no media and quasihydrostatic (NaCl) conditions, respectively. While in our case, solid quasihydrostatic environment (steatite) was used. It is possible that hydrostatic compression may be more effective in compressing the $a$ parameter, but our structural measurements are hydrostatic and the T-O transition in both of our measurements, hydrostatic and non-hydrostatic, coincides.

Finally, from our structural data, we show that the interlayer $Se(1)-Se(2)$ distance shows more important pressure dependence that the one reported in Ref 23. The $Se-Fe-Se$ angles and the Se height (not shown) show also a different behavior with pressure.

Thus, we believe that the mentioned discrepancies can be assigned to a small difference in Se stoichiometry that could reduce the superconducting critical temperature at ambient pressure of our sample, and induce a different structural and superconducting behavior under pressure.

In summary, we have studied the structural and superconducting properties of $FeSe$ under pressures up to $26 GPa$ using synchrotron radiation X ray diffraction and diamond anvil cells. We show that under compression, *FeSe* transforms from the original tetragonal structure to an orthorhombic one that has a higher superconducting temperature, $T_c > 34K$.

Acknowledgement

We wish to thank A. HADJ-AZZEM for the sample preparation.

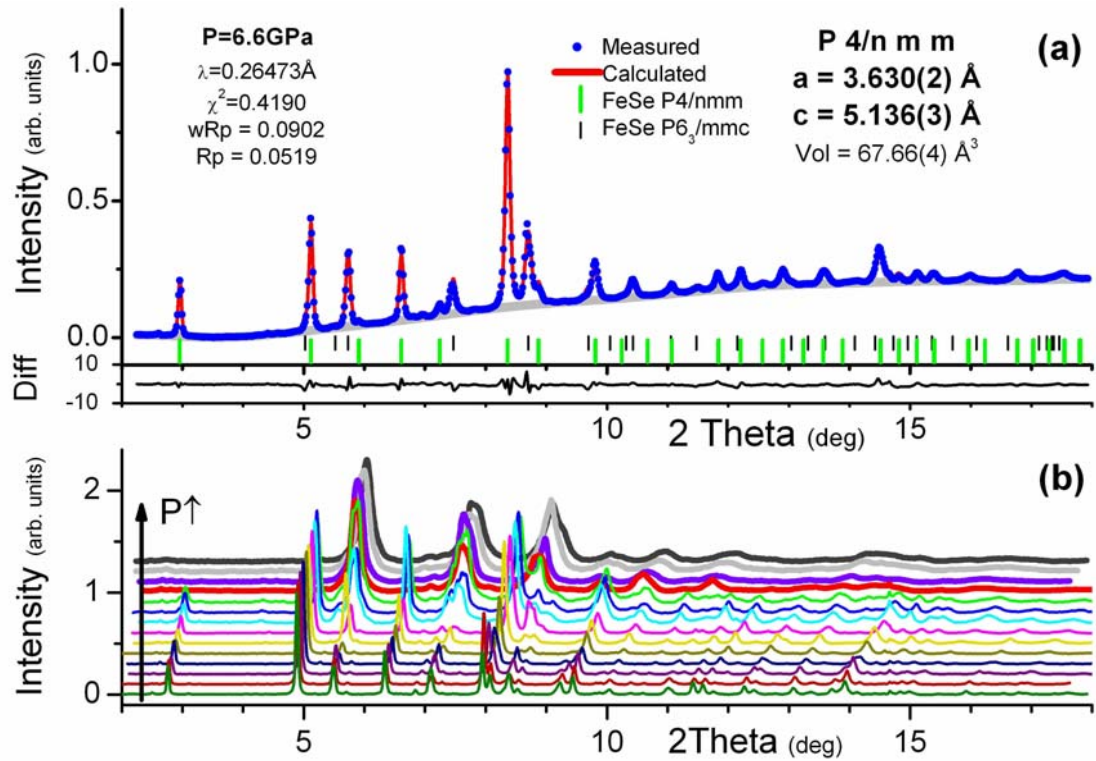

**Figure1:** (color online) **(a)** X-ray synchrotron radiation diffraction pattern of the *FeSe* powder sample at 6.6GPa (blue dots). The Rietveld refinement is the red solid line. The red lines correspond to the reflections associated with the tetragonal phase with space group P4/nmm, while the black lines represent the reflections of the P6$_3$/mmc hexagonal phase. **(b)** Pressure evolution of the diffraction patterns of *FeSe*. The data correspond to 0, 0.4, 1.3, 2.5, 3.8, 5.3, 7.6, 9.5, 11.2, 12.2, 13.4, 15.4, 21.4 and 26.5GPa, respectively. The phase transition to the orthorhombic phase can clearly be observed above 12.5GPa (thick lines).

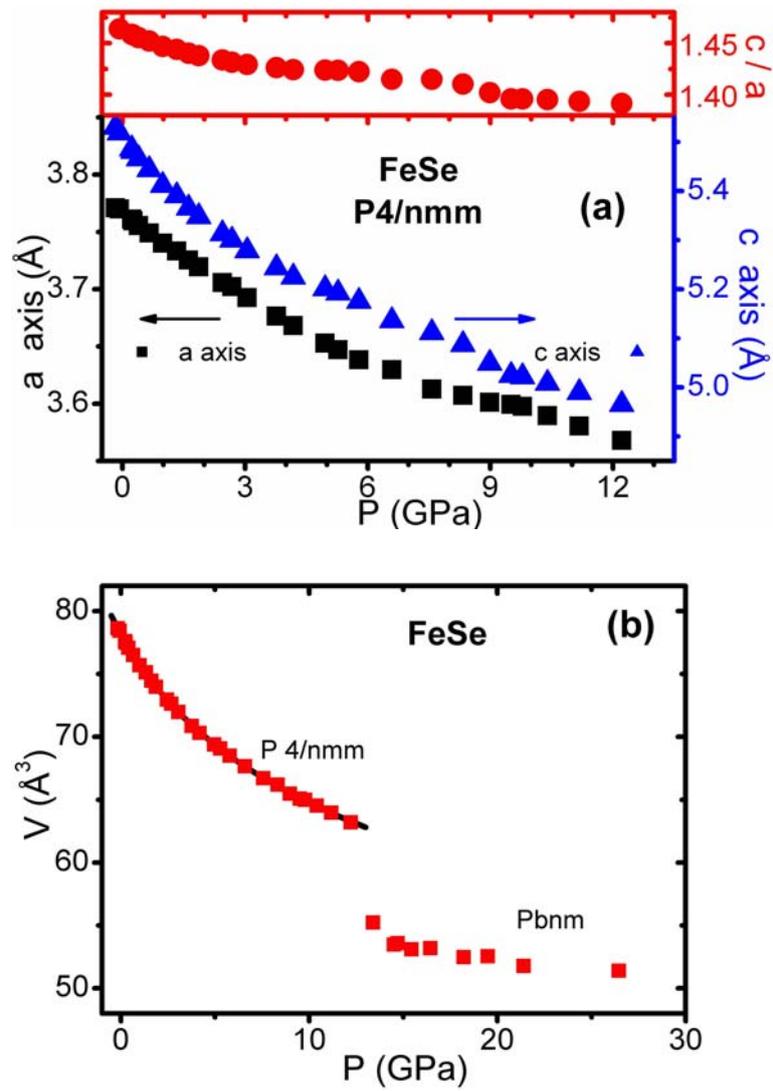

**Figure 2:** (color online) **(a)** Evolution of the lattice parameters and the $c/a$ ratio of the *FeSe* tetragonal phase as a function of pressure. Black squares: $a$; blue up triangles: $c$. **(b)** Pressure evolution of the volume of the tetragonal and the orthorhombic *FeSe* phases. The solid black line corresponds to the Murnaghan equation of state (see text).

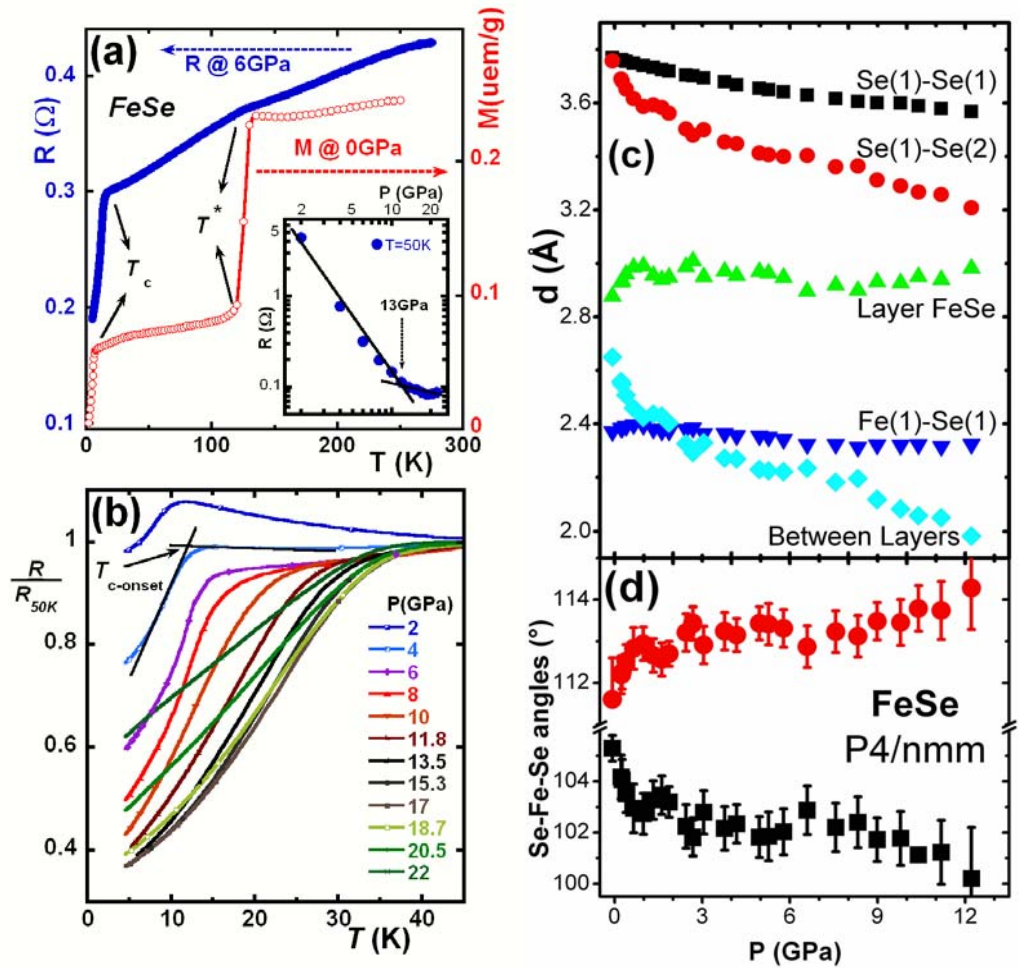

**Figure 3: (a):** Magnetization of the powder sample at ambient pressure showing the anomaly at $T^*$ and the superconducting transition temperature $T_c$. **Inset:** Pressure dependence of the resistance at 50K. It can be observed the kink at 13GPa. The lines are guide to the eye. **(b):** Electrical resistance of the *FeSe* sample as a function of temperature normalized to its value at 50K for different pressures. The definition of $T_{c\text{-onset}}$ is shown. Notice that the highest value is attained at 17GPa and that choosing other definitions of $T_{c\text{-onset}}$ can yield even higher values. **(c):** Pressure dependence of: the intralayer ($Se(1) - Se(1)$) and interlayer ($Se(1) - Se(2)$) Se atoms distance, of the Fe atoms distance ($Fe(1) - Fe(1)$), of the *FeSe* layer thickness (*Layer FeSe*) and of the distance between two *FeSe* layers (Between Layers) along c axis. **(d):** Pressure dependence of the angles $Se - Fe - Se$ in the tetragonal phase.

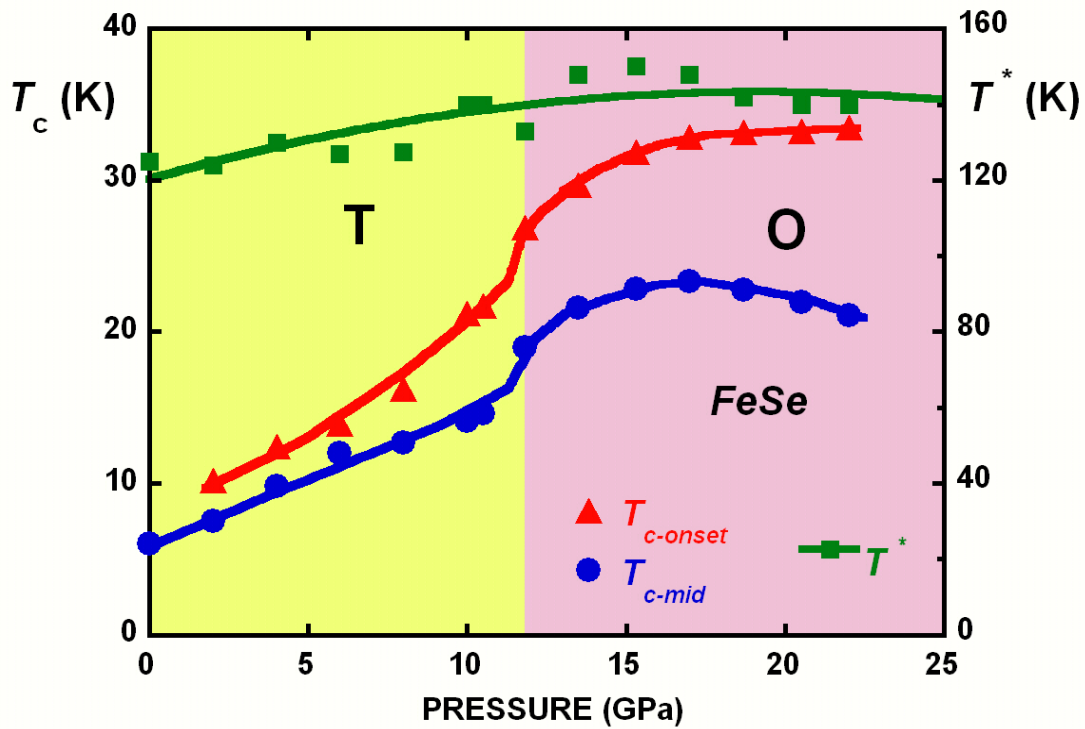

**Figure 4:** (color online) Evolution of the transition temperatures as a function of pressure. We observe that the transition at $T^*$ varies slightly with pressure and does not seem to be sensitive to the phase transition. In contrast, we do observe a change of behavior when we enter into the orthorhombic phase at 12.5GPa. The behavior of the middle of the superconducting transition follows a typical parabolic dependence with a maximum at around 16GPa, while the onset of the superconducting transition continues to increase up to our highest pressure, reaching 34K. The lines are guide to the eye.

---

[1] KAMIHARA Y., WATANABE T., HIRANO M. and H. HOSONO, *J. Am. Chem. Soc*. **130** (2008) 3296.

[2] TAKAHASHI H., IGAWA K., ARII K., KAMIHARA Y., HIRANO M. and HOSONO H., Nature (London) **453** (2008) 376.